# Criteria on quantum fluctuations of vacuum and photons influence in Spontaneous-Parametric Down-Conversion and Four-Wave-Mixing

Benoît Boulanger[1,*], Gaspar Mougin-Trichon[1], and Véronique Boutou [1]

[1] Université Grenoble Alpes, CNRS, Grenoble INP, Institut Néel, 38000 Grenoble, France

[*] benoit.boulanger@neel.cnrs.fr

**Abstract –** This article is a theoretical and quantitative exploration of the limit regarding the pump intensity between the two regimes of spontaneous-parametric down-conversion (SPDC) as well as of four-wave-mixing (FWM) in the framework of a semi-classical model and analytical calculations. A dimensionless parameter has been defined at this limit, corresponding to the photon-pairs flux *per* frequency unit: it has been found equal to 0.369. The ratio between the electric field of the generated photons and the quantum fluctuations of vacuum calculated at the limit is equal to 1.718. These quantitative results confirm that below the limit, the pump photon splitting leading to photon-pairs can be considered as spontaneous, *i.e.* mainly seeded by the quantum fluctuations of vacuum, while it is stimulated by the generated signal and idler photons above the limit, which corresponds to an optical parametric amplification regime. Our calculations also show that this limit can be easily reached in the case of SPDC according to the typical values of non-linearities and available crystal dimensions. In the case of FWM, it would be only possible in kilometric optical fibers. This corpus is a useful tool box for designing further quantum experiments performed from either side of the limit, as well as at the limit it-self where the influence of the quantum fluctuations of vacuum and of the generated photons should have the same weight. Furthermore, the quantum significance of the numerical values of the two criteria defined here remains to be established, which should motivate future theoretical quantum studies.

## 1 Introduction

Spontaneous-parametric down-conversion (SPDC) is the well-known and emblematic process of nonlinear optics leading to the generation of coherent tunable light and photon-pairs [1,2]. It is well known that the quantum properties of the generated photon-pairs vary as a function of the pump intensity. For instance, entangled photons-pairs sources are the most performant at low pump intensities, while high pump intensities lead to higher Einstein-Podolsky-Rosen (EPR) squeezing values [3,4]. A substantial body of literature investigates the quantum properties of SPDC sources, particularly in pulsed regimes, focusing on how pair correlations and entanglement evolve as a function of pump intensity [5] or source brightness [6]. Correlatively, it has been identified two pump intensity regimes regarding the flux of generated photon-pairs: one linear at low pump intensity, the other exponential at high pump intensity [3,4,7]. The limit between the two regimes has been experimentally found at a pump intensity of 40 MW/cm² at 532 nm when using a type II phase-matched KTP crystal [7].

The present work is not devoted to the quantum properties of parametric light but on its generation, with the goal to define and to calculate a dimensionless parameter independent of the nonlinear medium delimiting the two regimes of photon-pairs generation mentioned above, as well as to compute the associated ratio between the electric field of the generated signal and idler photons and the quantum fluctuations of the vacuum. The case of four-wave-mixing (FWM) is also considered since it is formally identical to SPDC from the modelling point of view [8,9]. Our motivation is to bring a quantitative framework for the design of further quantum measurements at the limit between the two regimes of photon-pairs generation as well as from either side.

The photonic diagrams of SPDC and FWM are reminded Fig. 1. Note that SPDC is governed by the second-order electric susceptibility $\chi^{(2)}$, while FWM depends on the third-order electric susceptibility $\chi^{(3)}$, both of these two elementary processes leading to the birth of photon-pairs.

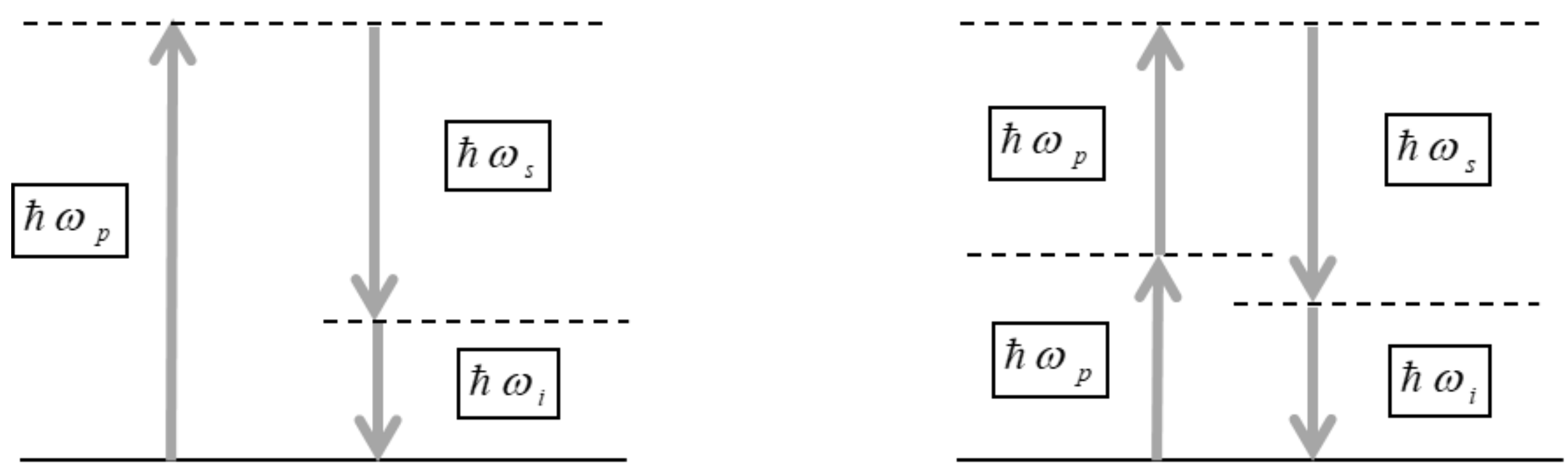

**Figure 1:** photonic diagrams of Spontaneous-Parametric Down-Conversion ($\chi^{(2)}: \omega_p \rightarrow \omega_s + \omega_i$) and Four-Wave-Mixing ($\chi^{(3)}: \omega_p + \omega_p \rightarrow \omega_s + \omega_i$) where $\chi^{(2)}$ and $\chi^{(3)}$ are the second-order and third-order electric susceptibilities, while *p*, *s* and *i* stand for pump, signal and idler, respectively.

## 2 Magnitude of the photon-pairs flux *per* frequency unit

The starting point of the present study is the following equation relative to a collinear phase-matched SPDC that has been recently proposed in the framework of a semi-classical model and experimentally validated between 1 W/cm$^2$ to 10 GW/cm$^2$ of pump intensity at 532 nm in a type II phase-matched bulk KTP crystal under the undepleted pump approximation (UPA), *i.e.* [7]:

$$N_{\text{pairs}}(L)=\frac{\varepsilon_0 n_s cS}{4\hbar\omega_s}\left[\left\|\Delta E_s^{\text{vacuum}}\right\|\left(\cos h(\beta L)-1\right)+\sqrt{\frac{\omega_s n_i}{\omega_i n_s}}\left\|\Delta E_i^{\text{vacuum}}\right\|\sin h(\beta L)\right]^2 \tag{1}$$

with

$$\begin{cases}\beta=\chi_{eff}^{(2)}\left\|E_p(0)\right\|\sqrt{\kappa_s\kappa_i}\\ \kappa_{s,i}=\dfrac{\omega_{s,i}}{2n_{s,i}c}\end{cases} \tag{2}$$

$L$ is the interaction length, $S$ is the section of the interaction path, *i.e.* corresponding to the overlap between the three interacting beams, $n_{s,i}$ is the refractive index at the circular frequency $\omega_{s,i}$, $\left\|E_p(0)\right\|$ is the modulus of the complex pump electric field amplitude, $\chi_{eff}^{(2)}$ is the second-order electric susceptibility and $\left\|\Delta E_{s,i}^{\text{vacuum}}\right\|$ corresponds to the quantum fluctuations of vacuum expressed as:

$$\left\|\Delta E_{s,i}^{\text{vacuum}}\right\|=\sqrt{\frac{\hbar\omega_{s,i}}{4\pi c\varepsilon_0 n_{s,i}S}\Delta\omega_{s,i}} \tag{3}$$

The quantity $\Delta\omega_{s,i}$ corresponds to the spectral linewidth of the generated photon-pairs.

By considering that $\Delta\omega_s=\Delta\omega_i(\equiv\Delta\omega)$ assuming a monochromatic pump wave, *i.e.* $\Delta\omega_p=0$, and inserting Eq. (3) in Eq. (1), it comes:

$$N_{\text{pairs}}(L)=\frac{\Delta\omega}{16\pi}\left[\left(\cosh(\beta L)-1\right)+\sinh(\beta L)\right]^2 \tag{4}$$

Since $\cosh(\beta L)+\sinh(\beta L)=\exp(\beta L)$, Eq. (4) reduces to:

$$N_{\text{pairs}}(L)=\frac{\Delta\omega}{16\pi}\left[\exp(\beta L)-1\right]^2=\frac{\Delta\nu}{8}\left[\exp(\beta L)-1\right]^2 \qquad (5)$$

It is now possible to calculate the flux of pairs *per* frequency unit, *i.e.*:

$$\left(\frac{N_{\text{pairs}}(L)}{\Delta\nu}\right)=\frac{1}{8}\left[\exp(\beta L)-1\right]^2 \qquad (6)$$

Note that $\left(\frac{N_{\text{pairs}}(L)}{\Delta\nu}\right)$ has no unit, it is a number of pairs.

According to the similarity of the coupled wave systems $\frac{\partial E_{p,s,i}(Z)}{\partial Z}$ of SPDC and FWM [8,9], the calculations in the case of FWM are the same than the previous ones by simply replacing $\beta$ in Eq. (6) by:

$$\beta=\frac{1}{2}\chi_{eff}^{(3)}\left\|E_p(0)\right\|^2\sqrt{\kappa_s\kappa_i} \qquad (7)$$

where $\left\|E_p(0)\right\|$ is the total pump amplitude, *i.e.* corresponding to the two pump waves.

### 3 Features of the limit between the two regimes of photon-pairs generation

Equation (6) exhibits a quadratic part and an exponential part as a function of $(\beta L)$, *i.e.*:

$$
\begin{cases}
\left(\dfrac{N_{\text{pairs}}(L)}{\Delta\nu}\right) = \dfrac{1}{8}\left[\beta L\right]^2 & (\beta L << 1) \\
\left(\dfrac{N_{\text{pairs}}(L)}{\Delta\nu}\right) = \dfrac{1}{8}\exp(2\beta L) & (\beta L >> 1)
\end{cases}
\tag{8}
$$

$(\beta L << 1)$ is called the small-signal regime in the following, while $(\beta L >> 1)$ corresponds to the high-signal regime.

Note that since $\left(\dfrac{N_{\text{pairs}}(L)}{\Delta\nu}\right)$ is proportional to $\beta^2$ in the small-signal regime according to Eq. (8) where $\beta$ is proportional to $\left\|E_p(0)\right\|$ in the case of SPDC (*cf.* Eq. (2)) and to $\left\|E_p(0)\right\|^2$ in the case of FWM (*cf.* Eq. (7)). This means that the evolution of $\left(\dfrac{N_{\text{pairs}}(L)}{\Delta\nu}\right)$ with the pump intensity in the small-signal regime is linear for SPDC while it is quadratic in the case of FWM. However, in the high-signal regime, the evolution is exponential in both cases.

Then, in order to describe both SPDC and FWM on the same graph, $\left(\dfrac{N_{\text{pairs}}(L)}{\Delta\nu}\right)$ is plotted in Fig. 2 as a function of $(\beta L)$ since they exhibit the same behavior as a function of this parameter.

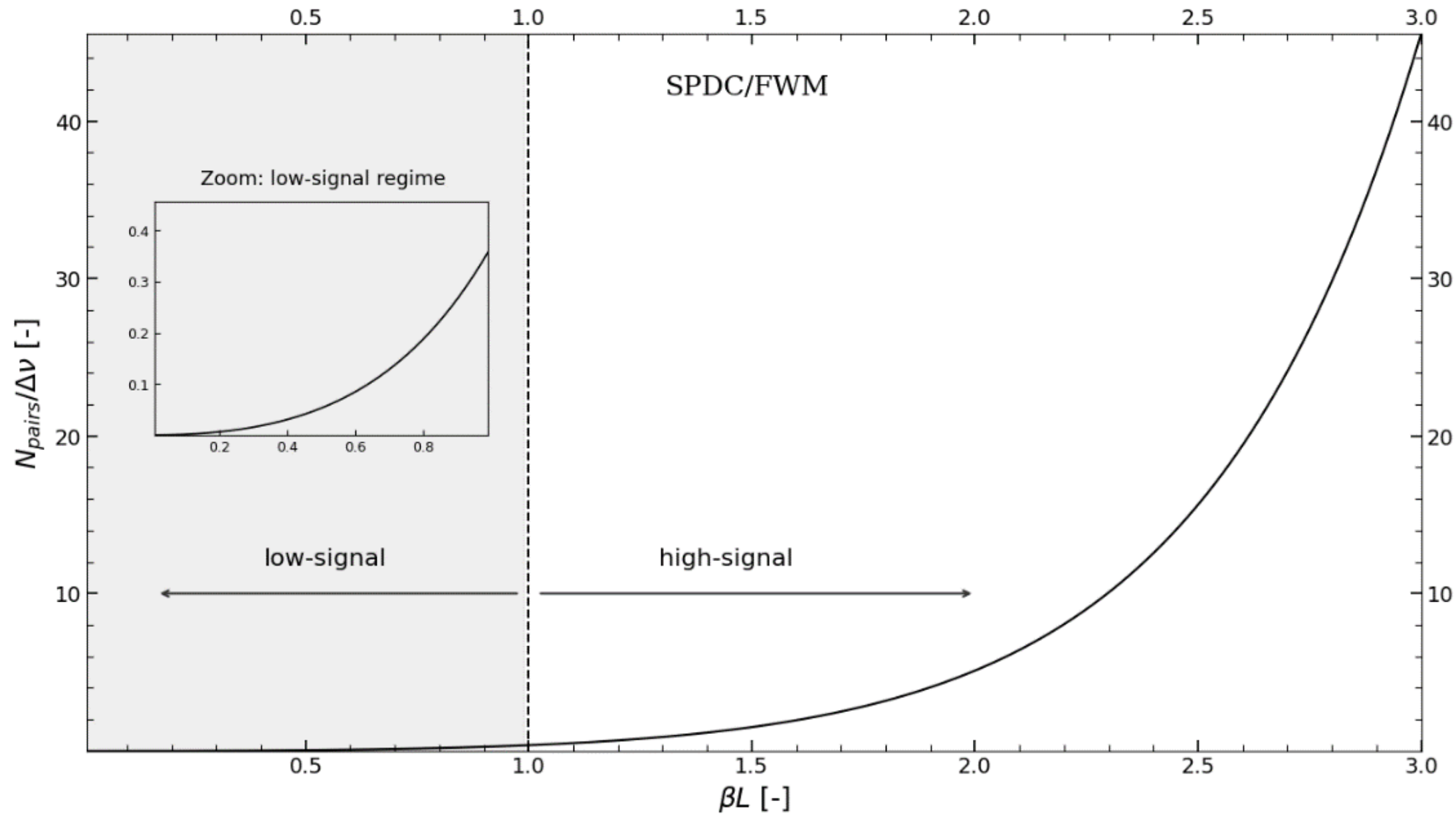

**Figure 2:** Photon-pairs flux *per* frequency unit $\left(\frac{N_{\text{pairs}}(L)}{\Delta\nu}\right)$ as a function of $(\beta L)$ in the cases of spontaneous-parametric down-conversion (SPDC) and four-wave-mixing (FWM). The dashed vertical line at $\beta L = 1$ corresponds to the limit between the weak signal regime (gray area) and strong signal regime (white area).

The limit between the two regimes is $\beta L = 1$. Then it comes for the pairs flux *per* frequency unit at the limit from Eq. (5):

$$\left(\frac{N_{\text{pairs}}}{\Delta\nu}\right)^{\text{lim}} = \frac{1}{8}\left[\text{e}-1\right]^2 = 0.369 \qquad (9)$$

Thus for $\beta L < 1$ there is less than 0.369 pair *per* second and *per* frequency unit, while it is more than 0.369 for $\beta L > 1$. Note that $\left(\frac{N_{\text{pairs}}}{\Delta\nu}\right)^{\text{lim}}$ does not depend on the interacting length *L* nor on the pump intensity $I_p(0)$. It is a "universal" limit between the two regimes.

Note that the limit could also be expressed in term of signal and idler photons, *i.e.*:

$$\left(\frac{N_{s+i}}{\Delta\nu}\right)^{\text{lim}} = 2\left(\frac{N_{\text{pairs}}}{\Delta\nu}\right)^{\text{lim}} = 0.738 \simeq 1 \tag{10}$$

It is now easy to compare the magnitudes of the electric field of the generated signal or idler photons and that of vacuum at the limit. Concerning the photons, knowing that $N_{s,i} = N_{\text{pairs}}$ and $N_{s,i} = \frac{\varepsilon_0 n_{s,i} c S}{4h\nu_{s,i}} \left\| E_{s,i} \right\|^2$, it comes:

$$\left\| E_{s,i} \right\|^{\text{lim}} = (e-1)\sqrt{\frac{h\nu_{s,i}}{2c\varepsilon_0 n_{s,i} S}\Delta\nu_{s,i}} \tag{11}$$

Concerning the vacuum, it comes from Eq. (3):

$$\left\| \Delta E_{s,i}^{\text{vacuum}} \right\| = \sqrt{\frac{h\nu_{s,i}}{2c\varepsilon_0 n_{s,i} S}\Delta\nu_{s,i}} \tag{12}$$

Then the ratio between Eq. (11) and Eq. (12) leads to:

$$\frac{\left\| E_{s,i} \right\|^{\text{lim}}}{\left\| \Delta E_{s,i}^{\text{vacuum}} \right\|} = (e-1) = 1.718 \simeq 1 \tag{13}$$

Then Eq. (10) and Eq. (13) allow us to establish that, when $\left\|E_{s,i}\right\| < \left\|\Delta E_{s,i}^{\text{vacuum}}\right\|$, which corresponds to $\left(\frac{N_{s+i}}{\Delta \nu}\right) < 1$, the SPDC or FWM efficiency is proportional to $(\beta L)^2$ defining the small-signal regime ; and when $\left\|E_{s,i}\right\| > \left\|\Delta E_{s,i}^{\text{vacuum}}\right\|$, which corresponds to $\left(\frac{N_{s+i}}{\Delta \nu}\right) > 1$, the SPDC or FWM efficiency is proportional to $\exp(2\beta L)$ defining the high-signal regime.

From a practical point of view, it can be useful to calculate the pump intensity corresponding to the limit, $I_p^{\lim}$, between the two regimes. It would be interesting for example to perform quantum measurements exactly at this limit where the influence of the quantum fluctuations of vacuum and of the generated photons should have the same weight. This calculation of $I_p^{\lim}$ can be done from Eq. (2) in the case of SPDC and Eq. (7) for FWM by stating $\beta L = 1$ and using $I = \frac{1}{2}\frac{n}{c\mu_0}\left\|E\right\|^2$. Then it comes in the case of SPDC:

$$I_p^{\lim/SPDC} = \frac{1}{2\pi^2 \mu_0 c} \frac{n_p n_s n_i \lambda_s \lambda_i}{\left[L\chi_{eff}^{(2)}\right]^2} \quad (13)$$

Figure 3 shows the evolution of the effective limit pump intensity $\Gamma^{SPDC}$, defined as $\Gamma^{SPDC} = \frac{I_p^{\lim/SPDC}}{n_p n_s n_i}$, as a function of $L$ for $\lambda_s = \lambda_i = 1\mu m$ and for different values of $\chi_{eff}^{(2)}$.

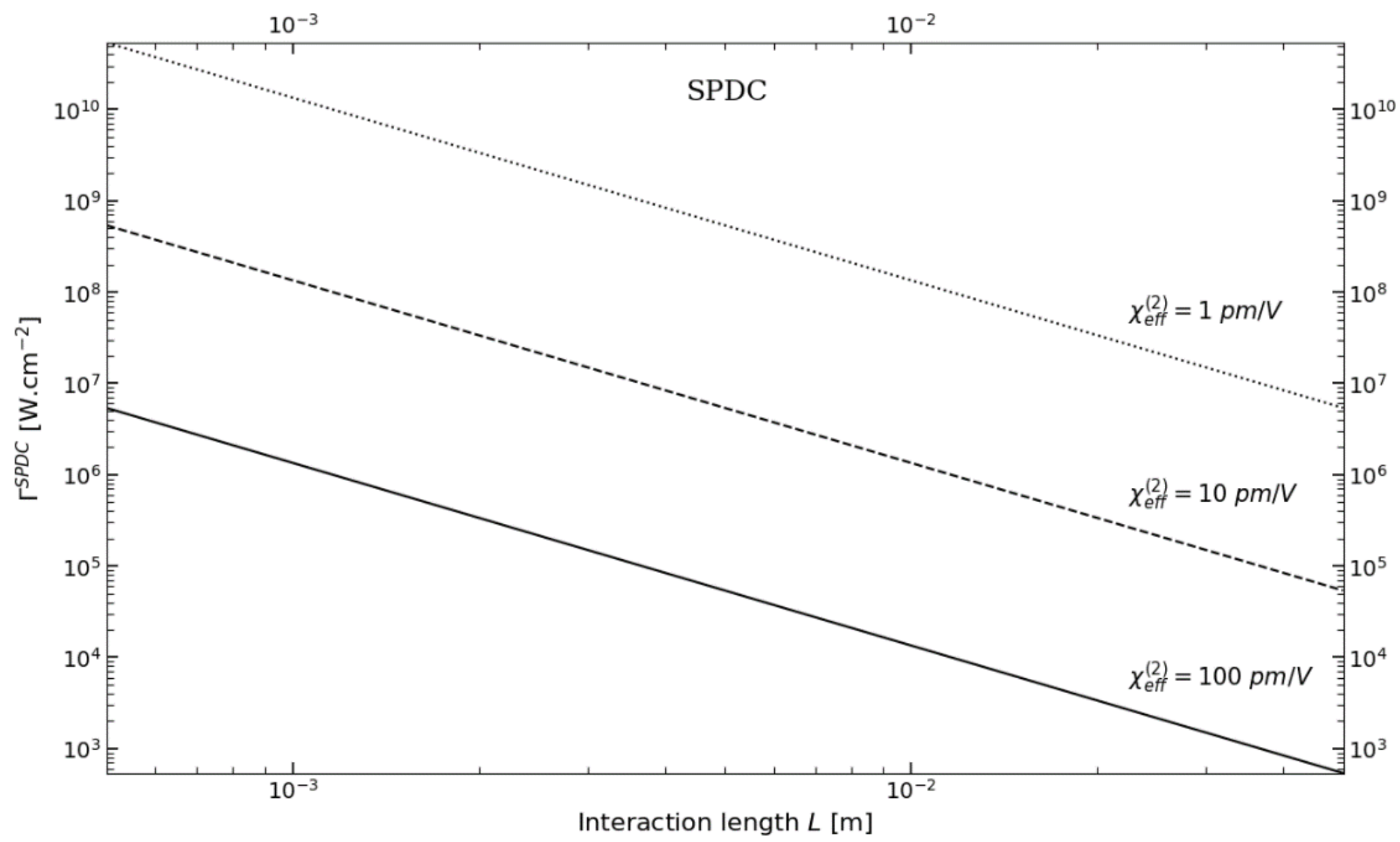

**Figure 3 :** calculated effective limit pump intensity of spontaneous-parametric down-conversion (SPDC) $\Gamma^{SPDC}$ as a function of the interaction length $L$ for $\lambda_s = \lambda_i = 1\mu m$ and three values of the second-order effective coefficient $\chi^{(2)}_{eff} = 1-10-100\, pm/V$.

Figure 3 shows that $\Gamma^{SPDC}$ ranges between 13.5 GW/cm$^2$ for $L = 1mm$ to 135 MW/cm$^2$ for $L = 1cm$ when $\chi^{(2)}_{eff} = 1pm/V$. This range is 135 MW/cm$^2$ – 1.35 MW/cm$^2$ for $\chi^{(2)}_{eff} = 10\, pm/V$ and 1.35 MW/cm$^2$ – 13.5 kW/cm$^2$ for $\chi^{(2)}_{eff} = 100\, pm/V$. Then $I_p^{\lim}$ can be easily reached according to the typical values of non-linearities and possible crystal lengths $(1mm < L < 1m)$ as in the cases of KTP and BBO $(\chi^{(2)}_{eff} \approx 1pm/V)$ [7,10], PPKTP and PPLN $(\chi^{(2)}_{eff} \approx 10\, pm/V)$ [11,12] or CSP and GaAs $(\chi^{(2)}_{eff} \approx 100\, pm/V)$ [13,14].

In the case of FWM, the limit pump intensity is expressed as:

$$I_p^{\lim/FWM} = \frac{1}{\pi}\sqrt{\frac{\varepsilon_0}{\mu_0}}\,\frac{n_p\sqrt{n_s n_i \lambda_s \lambda_i}}{L\chi^{(3)}_{eff}} \tag{14}$$

Figure 4 shows the evolution of the effective limit pump intensity $\Gamma^{FWM} = \dfrac{I_p^{\lim/FWM}}{n_p\sqrt{n_s n_i}}$ as a function of $L$ for $\lambda_s = \lambda_i = 1\mu m$ and for different values of $\chi_{eff}^{(3)}$.

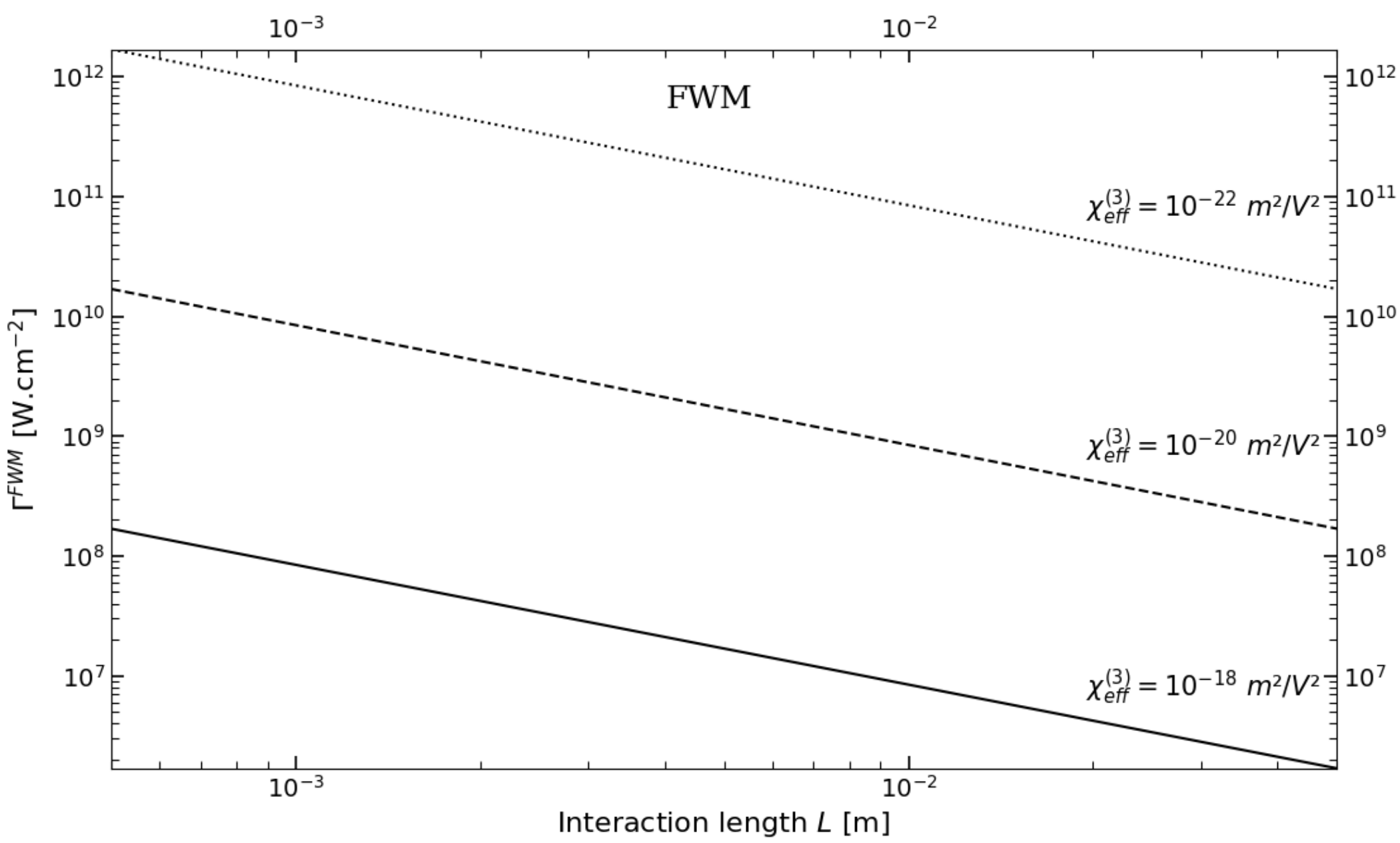

**Figure 4 :** calculated effective limit pump intensity of four-wave-mixing (FWM) $\Gamma^{FWM}$ as a function of the interaction length $L$ for $\lambda_s = \lambda_i = 1\mu m$ and three values of the second-order effective coefficient $\chi_{eff}^{(3)} = 10^{-22} - 10^{-20} - 10^{-18} m^2/V^2$.

Figure 4 indicates that $\Gamma^{FWM}$ ranges between 845 GW/cm$^2$ for $L = 1mm$ to 84.5 GW/cm$^2$ for $L = 1cm$ when $\chi_{eff}^{(3)} = 10^{-22} m^2/V^2$, this range being 8.45 GW/cm$^2$ – 845 MW/cm$^2$ for $\chi_{eff}^{(3)} = 10^{-20} m^2/V^2$ and 84.5 MW/cm$^2$ – 8.45 MW/cm$^2$ for $\chi_{eff}^{(3)} = 10^{-18} m^2/V^2$. The typical third-order non-linearity of crystals ranging between $10^{-22} m^2/V^2$ and $10^{-21} m^2/V^2$ [15], and the typical crystal lengths between 1 mm and 1 cm, it is obvious that FWM in such a media is always performed below $I_p^{\lim}$, that is to say in the small signal regime. Nevertheless, $I_p^{\lim}$ could be reached in optical fibers thanks to longer interaction lengths, as for example in the case of Silica fibers ($\chi_{eff}^{(3)} \approx 10^{-22} m^2 / V^2$) [16] for which $\Gamma^{FWM}$ is equal to 84.5 MW/cm$^2$ for $L = 10m$ and 845 kW/cm$^2$ for $L = 1km$.

## 4 Conclusion

The generation of photon-pairs has been theoretically investigated using a semiclassical approach where the quantum fluctuations of vacuum are taken as initial conditions of phase-matched spontaneous-parametric down-conversion (SPDC) and four-wave mixing (FWM) in the undepleted pump approximation (UPA).

The flux of photon-pairs $N_{\text{pairs}}$ expressed as a function of a quantity defined by the product of the nonlinearity and the interaction length, *i.e.* $\beta L$, exhibits two regimes: the so-called small-signal regime when $\beta L << 1$ leading to $N_{\text{pairs}} \propto (\beta L)^2$, and the strong-signal regime for $\beta L >> 1$ where $N_{\text{pairs}} \propto \exp(2\beta L)$.

We have characterized the limit, *i.e.* $\beta L = 1$, in term of a dimensionless quantity independent of the nonlinear medium that is defined as the photon-pairs flux *per* frequency unit, *i.e.* $\left( \frac{N_{\text{pairs}}}{\Delta \nu} \right)^{\lim}$. It has been found equal to 0.369, which gives for the generated signal (s) and idler (i) photons $\left( \frac{N_{s+i}}{\Delta \nu} \right)^{\lim} = 2 \left( \frac{N_{\text{pairs}}}{\Delta \nu} \right)^{\lim} = 0.738 \simeq 1$ photon *per* s and *per* Hz. We have also calculated at this limit the ratio between the magnitudes of the electric fields of the generated photons – signal or idler - and of vacuum $\frac{\left\| E_{s,i} \right\|^{\text{lim}}}{\left\| \Delta E_{s,i}^{\text{vacuum}} \right\|}$ that is equal to 1.718 ≈1.

The fact that these two criteria at the limit correspond to 1 is remarkable by its simplicity. However, the quantum significance of this feature remains to be established, which should motivate future theoretical quantum studies.

The relation of order regarding the electric fields allows us to confirm and well understand that the quantum fluctuations of vacuum governs SPDC as well as FWM in the small-signal regime, *i.e.* $\left\| E_{s,i} \right\| < \left\| \Delta E_{s,i}^{\text{vacuum}} \right\|$ meaning that the pump photons splitting into photon-pairs is a spontaneous process, while the splitting is stimulated by the signal and idler photons in the high-signal regime, *i.e.* $\left\| E_{s,i} \right\| > \left\| \Delta E_{s,i}^{\text{vacuum}} \right\|$, which corresponds to a regime of parametric amplification.

Then our study complements previous works reporting different quantum properties of the photon-pairs generated by SPDC as a function of the pump intensity, from entangled state of the signal and idler fields in the low-signal regime [17] to highly-squeezed EPR states in the high-signal regime [18]. Furthermore, it provides a unified description of SPDC and FWM.

The pump intensity corresponding to the limit has been also plotted as a function of the interaction length for different values of the second-order electric susceptibility in the case of SPDC and of the third-order electric susceptibility for FWM. Our calculations show that this limit can be easily reached in the case of SPDC according to the typical values of non-linearities and crystal dimensions; it is more difficult for FWM but possible, in particular in the case of optical fibers for which the interaction length can be several kilometers. Then, knowing quantitatively the pump intensity corresponding to the boundary between the two regimes of low and high signal regimes will be then a useful guide for further quantum calculations, and should provide new insights into revisited quantum measurements. Actually, it would be of prime interest to characterize this limit it-self, where the influence of the quantum fluctuations of vacuum and of the generated photons should have the same weight, by using pair-coincidence experiments similar to pioneer works [5,19,20].

**Author contributions:**

**BB** Conceptualization, Methodology, Calculation, Writing – original draft-review & editing, **GMT** Calculation, software, validation, writing-review, **VB** validation, writing-review.

**Acknowledgment**

The authors wish to thank Sebastien Tanzilli for fruitful discussions.

**Funding**

The PhD fellowship of Gaspar Mougin-Trichon is funded by the Program QuanTEdu-France n° ANR-22-CMAS-0001 France 2030.

**Conflicts of interest**

The authors have nothing to disclose.

**Data availability statement**

Data are available upon reasonable request at the indicated e-mail.